\documentclass[fleqn,usenatbib,twocolumn]{mnras}

\usepackage{newtxtext,newtxmath}

\usepackage[T1]{fontenc}

\DeclareRobustCommand{\VAN}[3]{#2}
\let\VANthebibliography\thebibliography
\def\thebibliography{\DeclareRobustCommand{\VAN}[3]{##3}\VANthebibliography}

\usepackage{graphicx}	
\usepackage{amsmath}	
\usepackage{setspace}
\usepackage{caption}
\usepackage{subcaption}
\usepackage{listings}
\usepackage{color}
\usepackage{float}
\usepackage{widetext}

\usepackage[normalem]{ulem}

\newcommand{\cii}{[\ion{C}{ii}]}
\newcommand{\myr}{\rm M_{\odot}yr^{-1}}
\newcommand{\xp}{x_{\parallel}}
\newcommand{\xpa}{x_{\parallel,1}}
\newcommand{\xpb}{x_{\parallel,2}}
\newcommand{\xperp}{\boldsymbol{x}_{\perp}}
\newcommand{\kp}{k_{\parallel}}
\newcommand{\kpa}{k_{\parallel,1}}
\newcommand{\kpb}{k_{\parallel,2}}

\newcommand{\thetav}{\boldsymbol{\theta}} 
\newcommand{\ellv}{\boldsymbol{\ell}}
\newcommand{\VEV}[1]{\left\langle #1 \right\rangle}

\newcommand{\SED}{S_{\nu}^{\rm eff}}
\newcommand{\g}{\text{g}} 
\newcommand{\cov}{\text{Cov}}

\title[Joint CIB and $\cii$ SFR Constraint]{Constraining the Star Formation Rate using Joint CIB Continuum and $\cii$ Intensity Mapping}

\author[Zhou \& Maniyar \& Pullen]{
Zilu Zhou,$^{1}$\thanks{E-mail: zz1994@nyu.edu}
Abhishek S. Maniyar,$^{1}$
Anthony R. Pullen$^{1, 2}$
\\
$^{1}$Department of Physics, New York University, 726 Broadway, New York, NY, 10003, U.S.A.\\
$^{2}$Center for Computational Astrophysics, Flatiron Institute, New York, NY 10010, U.S.A.}

\date{Accepted XXX. Received YYY; in original form ZZZ}

\pubyear{2022}

\begin{document}
\label{firstpage}
\pagerange{\pageref{firstpage}--\pageref{lastpage}}
\maketitle

\begin{abstract}
Line intensity mapping (LIM) experiments probing the nearby universe can expect a considerable amount of cosmic infrared background (CIB) contiuum emission coming from near and far-infrared galaxies. For the purpose of using the LIM data to constrain the star formation rate (SFR), we argue that the CIB continuum - traditionally treated as contamination - can be combined with the LIM signal to enhance the SFR constraints achievable. We first present a power spectrum model that is capable of joining continuum and line emissions that assume the same prior SFR model. We subsequently analyze the effectiveness of the joint model in the context of the EXperiment for Cryogenic Large-Aperture Intensity Mapping (EXCLAIM), which utilizes the $\cii$ molecular line to study the SFR. We numerically compute the theoretical power spectra according to our model and the EXCLAIM survey specifics, and perform Fisher analysis to obtain SFR parameter constraints. We find that although the joint model has no considerable advantage over LIM alone assuming the current survey level of EXCLAIM, its effects become significant when we consider more optimistic values of survey resolution and angular span that are expected of future LIM experiments. By manipulating the Fisher formalism, we show that the CIB is not only an additional SFR sensitive signal, but also serves to break the SFR parameter degeneracy that naturally emerges from the $\cii$ Fisher matrix. For this reason, addition of the CIB will allow improvements in the survey parameters to be better reflected in the SFR constraints, and can be effectively utilized by future LIM experiments.
\end{abstract}




\section{Introduction}

The star formation history of galaxies is a powerful reference for the study of galaxy evolution, composition, and the underlying large-scale structure (LSS) behavior \citep{madau14}. Its characterizing function, the star formation rate (SFR), has been probed extensively in recent years using observables with well modeled luminosity dependences on the SFR \citep{kenni12}.

Among these observables are the atomic and molecular spectral line emissions, detectable using the line intensity mapping (LIM) technique. In recent years, LIM has often been proposed as the preferred method for probing high redshift galaxies \citep{2010JCAP...11..016V,2011JCAP...08..010V,2017arXiv170909066K,2019BAAS...51c.101K}. Compared to the traditional point source resolution surveys, LIM maps the integrated spectral line luminosity in regions across the sky, with each galaxy contributing its brightness, thus bypassing the loss of resolution at high redshifts. Popular emission lines for LIM include the 21cm $\ion{H}{i}$ emission, the Ly-$\alpha$ emission, as well as the CO \& $\cii$ emission lines. There exist various models of the CO \citep{2008A&A...489..489R,2011ApJ...741...70L} and $\cii$ \citep{2015ApJ...806..209S,2015ApJ...813...36V,2019MNRAS.488.3014P} luminosity-SFR relation, making them excellent choices to conduct analyses of the SFR.

High-redshift LIM surveys are often subject to sources of foreground contamination. One particularly notable contamination is the thermal dust emission emitted from both the Milky Way (MW) and distant galaxies, which produces signals in the infrared (IR) that overlaps with many of the line emission frequencies, including CO and $\cii$. In particular, the collective, diffuse dust emissions from distant galaxies make up the \emph{cosmic infrared background} (CIB) \citep{1996A&A...308L...5P,2005ARA&A..43..727L}. The anisotropic CIB has been observed by Planck and Herschel at multiple frequencies across a wide range of angular scales \citep{2013ApJ...779...32V,2016A&A...596A.109P,2019ApJ...881...96V,2019ApJ...883...75L}. These continuum foregrounds are often highly dominant over the target emission lines and therefore pose a significant challenge to LIM technique. A key feature of the CIB is their smooth spectral energy distributions (SED) in Fourier space; consequently, continuum foregrounds reside in modes perpendicular to the line of sight (LOS), i.e. with $k_\parallel \simeq 0$. This fact renders these modes unusable to detect the target line emissions through LIM. At the same time, this feature also justifies a simple method of removing the foregrounds, by discarding any signal within these low $k_\parallel$ modes. For example, \citet{2019ApJ...872...82S} forecast a LOS power spectrum analysis comparing the CIB continuum and the $\cii$ line intensity, and finds that the CIB contaminates only the lowest two discrete $\kp$ modes, which is illustrated in Fig. 16 of \cite{2019ApJ...872...82S}. Therefore, by discarding these modes while using only the remaining ones, we can separate out the information contained in the target line while introducing only minimal bias.

While this cutoff treatment is straightforward and effective for general analyses of LIM, it treats CIB as a nuisance. As mentioned before, the CIB originates from galactic dust within the interstellar medium (ISM). These emissions are triggered when UV radiation from the formation of new stars are absorbed by the ISM dust, and re-emitted as IR radiation. As such, the emission of CIB is itself an excellent tracer of the star formation history \citep{maniyar18}; in fact, models have been developed that relate the CIB intensity to the SFR, such as \citet{maniyar21}. With this consideration, we argue that retaining the CIB part of the signal should not hinder the prospects of constraining the SFR, and may even be a beneficial procedure. Thus, with an appropriate choice of an SFR-dependent CIB model and an aforementioned spectral line luminosity-SFR relation, we can perform a joint analysis, including both the CIB continuum signal and the discrete line signal to turn the CIB "contamination" into an advantage. 

To analyze the effectiveness of this joint analysis, we apply our framework to the upcoming EXperiment for Cryogenic Large-Aperture Intensity Mapping (EXCLAIM) \citep{2021JATIS...7d4004S, 2020JLTP..199.1027A, 2020SPIE11445E..24C}. A LIM survey, EXCLAIM seeks to probe the star formation history in the near universe below $z \leq 3.5$ by observing the CO and $\cii$ lines. The $\cii$ line in particular serves as an excellent target line for this study, with numerous proposed and well studied $L_{\cii}-$SFR relations \citep{silva15, lagache18, chung20, schaerer20}. We choose the empirical relation given by \cite{yang21}, whose analysis also fixes EXCLAIM as a fiducial survey. Together with the CIB intensity model from \citet{maniyar21}, we use the joint intensities to forecast a $\cii$-CIB power spectrum analysis for the projected EXCLAIM $\cii$ observations within a redshift range of $2.5 \leq z \leq 3.5$.

The rest of this paper is organised as follows: in Sec.~\ref{sec:theory} we present our most general power spectrum model capable of combining the CIB continuum emission and an arbitrary line emission, with discussion of their vastly different behaviors over redshifts and how we resolve this difference. In Sec.~\ref{sec:models} we provide an overview of the models of the CIB and the $\cii$ line in relation to the SFR. We present numerical results of our power spectrum model in Sec.~\ref{sec:powerspec} for the specifics of the EXCLAIM survey. We use these power spectra results as inputs to Fisher analysis and present in Sec.~\ref{sec:fisher} the SFR constraint forecasts. We conclude our findings and discuss the applicability of our model to future LIM surveys in Sec.~\ref{sec:conclusion}. 
Where applicable, we assume the Planck 2018 best-fit cosmological parameters \citep{2020A&A...641A...6P} as the fiducial values.

\section{Bridging Continuum and Line Intensity Power Spectra} \label{sec:theory}
We begin by deriving our most general angular power spectrum model. 

\subsection{Redshift dependence \& Overdensities}
Combining intensities of continuum and line emissions is non-trivial due to their varying wavelength dependence on redshift. From the LIM perspective, a particular spectral line $X$ will have a determined rest frame wavelength $\lambda_X$. Familiar examples of this feature are the $\lambda_\ion{H}{i} = 21 \rm cm$ for $\ion{H}{i}$ and the $\lambda_{\cii} = 158 \micron$ for $\cii$ lines. Thus, we can draw a one-to-one correspondence between a line intensity signal received at some effective wavelength $\lambda$ and its redshift origin via the following relation:
\begin{equation}
    \lambda(z) = \lambda_X (1+z)\, .
    \label{redshift}
\end{equation}

On the other hand, continuum emissions like the CIB have no determined rest frame wavelength. Instead, at a given redshift, a continuum signal is emitted across a range of wavelengths according to its characteristic SED, often denoted as $\SED(z)$. Conversely, a continuum signal received at wavelength $\lambda$ cannot be uniquely traced to a particular redshift origin. This distinction directly affects how we define the intensity fluctuations of the line and continuum emissions, as we proceed towards a wavelength oriented power spectrum model.

We start with the definition of the line intensity fluctuation, owing to its aforementioned simple behavior over redshift. For a line X, we model this as
\begin{equation}
    \delta I_{X}(\xp', \thetav) = I_{X}(\xp') \times b_{X} \times \delta_m[\xp', \chi \thetav]\, ,
    \label{deltaline}
\end{equation}
where $b_X$ is the effective intensity bias, and $\delta_m$ is the underlying matter overdensity. Similar to \cite{2019ApJ...872...82S}, we divide our 3D coordinate system into parallel and perpendicular components with respect to the LOS, $\xp'$ and $\xperp'$. Assuming a small survey area, we adopt the flat-sky approximation and further decompose the perpendicular component as $\xperp' = \chi \thetav$, where $\chi$ is the comoving distance to the center of the LOS survey window, and $\thetav$ is the angular size of the transverse separation from the center. 

As is traditionally done, computing the power spectrum involves correlating the observable fluctuations in Fourier space. To proceed, we derive the analytical Fourier transform of the intensity fluctuation defined above as
\begin{align}
    \nonumber \delta I_X(\kp', \ellv') &= \int d\xp' d^2\thetav\ e^{i\kp'\xp'}e^{i\ellv'\cdot \thetav} \delta I_{X}(\xp', \thetav)\, ,\\
    \nonumber &= \int d\xp' d^2\thetav\ e^{i\kp'\xp'}e^{i\ellv'\cdot \thetav} I_{X}(\xp') b_{X}\\
    \nonumber &\times \frac{1}{\chi^2} \int \frac{d\kp d^2\ellv}{(2\pi)^3} e^{-i\kp\xp'} e^{-i\ellv \cdot \thetav} \delta_m\left(\kp, \frac{\ellv}{\chi}\right)\, ,\\
    &= \int \frac{d\xp' d\kp}{2\pi\chi^2} e^{i(\kp'-\kp)\xp'} I_X(\xp') b_X \delta_m\left(\kp, \frac{\ellv'}{\chi}\right)\, ,
    \label{linefourier}
\end{align}
where $\kp'$ and $\ellv'$ are the Fourier space duals of $\xp'$ and $\thetav$, respectively.  In the first line, the integration is defined over a finite volume, while we assume the angular region is much larger than the angular scales of interest such that we can approximate the angular integral to be over an infinite area. Furthermore, we express the matter overdensity introduced in Eq. \ref{deltaline} as the inverse Fourier transform of its Fourier counter part, $\delta_m(\kp, \frac{\ellv}{\chi})$, where we again use the flat-sky approximation to express the perpendicular wave mode as $\boldsymbol{k}_\perp = \frac{\ellv}{\chi}$. This substitution has the advantage of eliminating the angular part of the integral entirely, as the additional angular mode $\ellv$ produces a Dirac-delta function over $\ellv'-\ellv$. Note that the parallel part does not similarly produce a delta function, since for physical surveys the LOS distance will be over a finite window. While we can approximate this finite range via a sinc function, we choose to retain this general form as it is computationally feasible. 

The derivations so far are not dissimilar to the traditional treatment for line intensity. The challenge now lies in producing a similar, compatible expression for the CIB continuum. To begin with, the CIB intensity flucation can be written as follows
\begin{align}
    \delta I_{\rm CIB}(\xp', \thetav) = \int dz \frac{dI_{\rm CIB}(\xp')}{dz} \times b_{\rm CIB} \times \delta_m[\xp(z), \chi \thetav]\, ,
    \label{deltaCIB}
\end{align}
where $b_{\rm CIB}$ is the effective bias, and $I_{\rm CIB}$ is the CIB intensity model that depends on the underlying continuum SED. Here we introduce a new distance variable, $\xp$, which is the comoving distance to the CIB-emitting sources and must crucially be distinguished from $\xp'$, the distance scale we have worked with thus far. We use the primed coordinates to strictly represent the inferred distance based on the perceived wavelength of the signal (using Eq. \ref{redshift}), whether it is line or continuum in nature. Indeed, this treatment marks where we have "combined" line and continuum emissions. 

As we previously discussed, the primed coordinate sufficiently contains all the information of a line signal, whereas for continuum signal the primed coordinate does not uniquely fix its source redshift. Thus, we introduce the unprimed coordinates to control the redshift part of the CIB separately from its wavelength. To analyze the CIB emission at a particular wavelength, then, we perform an integral over the full range of the unprimed coordinate, where the intensity contribution to that wavelength at a certain redshift will be weighted by the SED within the intensity model. The unprimed coordinate also parameterizes the matter overdensity for the CIB fluctuations, as the underlying matter traces the physical origin of the CIB emission sources, and should not be associated with the perceived wavelengths by the detector.

Similarly as before, we perform a Fourier transform of the CIB flucation, which has the form
\begin{align} \label{fourierCIB}
    \nonumber \delta I_{\rm CIB}(\kp', \ellv') &= \int \frac{dz d\xp' d\kp}{2\pi\chi^2} e^{i[\kp'\xp'-\kp\xp(z)]} \frac{dI_{\rm CIB}(\xp')}{dz}\\ 
    &\times b_{\rm CIB} \delta_m\left(\kp, \frac{\ellv'}{\chi}\right)\, ,
\end{align}
thus obtaining the intensity fluctuations of the CIB and of the line emission that share the same set of Fourier space variables.

EXCLAIM-like surveys will rely on the cross-correlation of the LIM signals with galaxy surveys in order to separate the target line emission from the Milky-way and interloper emissions. To emulate this procedure, we will use as our observables the cross-correlation power spectra of the intensities with a galaxy overdensity field, within the redshift range of the line intensity origin. We can express this overdensity as
\begin{align}
    \delta \g(\xp', \thetav) = b_{\g} \delta_m[\xp', \chi\thetav]\, ,
    \label{eq:deltag}
\end{align}
where $b_{\g}$ is the galaxy bias. As this is simply an overdensity field, Eq. \ref{eq:deltag} does not contain any factors of brightness or intensity. We also keep the field coordinates consistent with that of the line intensity (primed coordinates), as the cross-correlation will pick up the LIM emission coming from the redshift range of the galaxy survey. Galaxies within the redshift range of the $\cii$ line for the EXCLAIM survey have been well surveyed by the eBOSS collaboration \citep{2016AJ....151...44D}, allowing for an effective cross-correlation procedure. Consistency of the coordinate system also leads to a Fourier space expression very similar to that of Eq. \ref{linefourier} for the line intensity
\begin{align}
    \delta \g(\kp', \ellv') = \int \frac{d\xp' d\kp}{2\pi\chi^2} e^{i(\kp'-\kp)\xp'} b_{\g} \delta_m\left(\kp, \frac{\ellv'}{\chi}\right)\, ,
    \label{eq:gfourier}
\end{align}
identical to Eq. \ref{linefourier} without the intensity part.

\subsection{Angular Power Spectrum}
Having defined all the relevant fluctuation and overdensity fields, we can derive the angular power spectra by modeling the cross-correlations of these quantities. The most general 3D power spectrum can be written as
\begin{align}
    \VEV{\delta X(\mathbf{k}_1) \delta Y^*(\mathbf{k}_2)} = (2\pi)^3 P^{X\times Y}(k_1)\delta^3{(\mathbf{k}_1-\mathbf{k}_2)} \, ,
    \label{eq:3dpower}
\end{align}
where $\delta^3(\mathbf{k}_1-\mathbf{k}_2)$  is the 3D Dirac-delta function and $P^{X\times Y}$ represents the 3D cross-correlation between observables $X$ and $Y$. Each $\mathbf{k}$ can subsequently be decomposed into their parallel and perpendicular modes $(\kp,\mathbf{k}_\perp)$ with $\mathbf{k}_\perp=\ellv/\chi$, and the angular power spectrum can be obtained by applying a flat-sky approximation. Eq. \ref{eq:3dpower} becomes

\begin{eqnarray}
    \nonumber \VEV{\delta X(\kpa', \ellv'_1) \delta Y^*(\kpb', \ellv'_2)} &=& (2\pi)^2 \mathcal{L}' C^{X\times Y}(\kpa',\ell'_1)\\
    &\times& \delta^2{(\ellv'_1-\ellv'_2)} \, ,
    \label{eq:3dangle}
\end{eqnarray}
where $\mathcal{L}'$ is the survey window size along the LOS (i.e. comoving distance span), and $C^{X\times Y}(\kp',\ell')$ is the angular cross-correlation power spectrum.  

The LHS of Eqs. \ref{eq:3dpower} \& \ref{eq:3dangle} represent the covariance of the observables overdensity fields. If $X$ and $Y$ are both line intensity signals, this operation can be expressed explicitly as
\begin{widetext}
\begin{eqnarray}
    \nonumber \VEV{\delta X(\kpa', \ellv'_1) \delta Y^*(\kpb', \ellv'_2)} &=& (2\pi)^3 \int \frac{d\kpa}{2\pi\chi^2}\frac{d\kpb}{2\pi\chi^2} d\xpa'd\xpb' \delta{(\kpa-\kpb)}\chi^2\delta^2{(\ellv'_1-\ellv'_2)}\\
    \nonumber &\times& \Bigg[b_XI_X e^{(\kpa'-\kpa)\xpa'}\Bigg]\Bigg[ b_YI_Y e^{(\kpb'-\kpb)\xpb'}\Bigg] \VEV{\delta_m\left(\kpa, \frac{\ellv'_1}{\chi(z_1)}\right) \delta_m\left(\kpb, \frac{\ellv'_2}{\chi(z_2)}\right)}\\
    \nonumber &=& (2\pi)^2 \int \frac{d\kp}{2\pi\chi^2} d\xpa'd\xpb' \delta^2{(\ellv'_1-\ellv'_2)} P\left(\sqrt{\kp^2 + \frac{\ell'^2}{\chi^2}}, z_1, z_2 \right)\\
    &\times& \Bigg[b_XI_X e^{(\kpa'-\kp)\xpa'}\Bigg]\Bigg[ b_YI_Y e^{(\kpb'-\kp)\xpb'}\Bigg]\, ,
    \label{eq:windowcov}
\end{eqnarray}
\end{widetext}
where $b_X$, $I_X$ and $b_Y$, $I_Y$ are bias and intensity components of signals $X$ and $Y$, respectively, and $P(k, z_1, z_2)$ is the matter power spectrum (MPS) coming from the covariance of the matter overdensities coupled to each signal. Similar expressions for cross-correlations of background intensity and galaxy field will differ only by the content of the integrand, and the exact components are specified by Eqs. \ref{fourierCIB} and \ref{eq:gfourier}, in the same way we used Eq. \ref{linefourier} in deriving Eq.\ref{eq:windowcov}.  

Here we must briefly discuss the MPS component in our model. Notice that in Eq.\ref{eq:windowcov}, we formulate the MPS to dependent on two redshift values, $z_1$, $z_2$, separately, each associated with one of the input observables to the cross-correlation. In most situations, if both matter overdensities are mapped over the same redshift range (i.e. both observables trace the same region of matter), this distinction is redundant. As we will see, however, the redshift ranges of the line and CIB intensities can vary quite drastically, depending on the fiducial survey. When cross-correlating two matter overdensities that span different redshift windows, it is not ideal to approximate the result as one MPS at a single redshift. Thus, we impose the redshift dependence of both observables such that the MPS correctly captures the correlation of matter over different ranges spanned by the observables.

Practically, the MPS with this consideration can be expressed in terms of the linear growth factor, $D(z)$, as
\begin{align}
    P(\kp, \ell', z_1, z_2) = P_0(\kp, \ell')D(z_1)D(z_2)\, ,
    \label{eq:linearmps}
\end{align}
where $P_0$ is the initial value of the MPS. This treatment is, however, an exact expression only when considering the linear MPS. Given that LIM will likely occur around low redshifts where non-linear corrections to the MPS are considerable, we should replace Eq. \ref{eq:linearmps} with the non-linear MPS. With this replacement, however, the redshift evolutions of two matter overdensities can no longer be trivially separated in terms of growth factors. To proceed, we rely on the approximation that the non-linear MPS correlated over two redshifts can be expressed as the geometric mean of its values at the two individual redshifts, namely
\begin{align}
    P(\kp, \ell', z_1, z_2) = \sqrt{P(\kp, \ell', z_1)} \times \sqrt{P(\kp, \ell', z_2)} \, .
    \label{eq:geomean}
\end{align}
We note that this treatment, when applied to the linear MPS, returns the exact result in Eq. \ref{eq:linearmps}, where taking the square root of each linear MPS simply extracts the corresponding linear growth factor. Since the non-linear MPS introduces only perturbative corrections to the linear version, Eq. \ref{eq:geomean} serves as an approximation even for the non-linear case.

With this method of separation, we can define observable window functions to drastically simplify the expressions for cross-correlation. To account for the redshift dependence on each square root factor of the non-linear MPS, we will directly couple the MPS to the window functions themselves, where they are integrated over the correct redshift ranges alongside other components of its "host" observable. According to the form of Eq. \ref{eq:windowcov} as well as the Fourier transformed overdensities of the observables we introduced (Eqs. \ref{linefourier}, \ref{fourierCIB}, and \ref{eq:gfourier}), the window functions can be written as
\begin{eqnarray}
    \nonumber \tilde{L}(\kp, \kp', \ell') &=& \int d\xp' e^{(\kp'-\kp)\xp'}\ I_X(\xp') b_X \\
    &\times& \sqrt{P\left(\sqrt{\kp^2 + \frac{\ell'^2}{\chi^2}}, z(\xp')\right)}\, ,
    \label{eq:lwindow}
    \\
    \nonumber \tilde{B}(\kp, \kp', \ell') &=& \int dz d\xp' e^{i(\kp'\xp'-\kp\xp(z))} \frac{dI_{\rm CIB}(\xp')}{dz}\\ 
    &\times& b_{\rm CIB}(z) \sqrt{P\left(\sqrt{\kp^2 + \frac{\ell'^2}{\chi^2}}, z\right)}\, , 
    \label{eq:bwindow}
    \\
    \nonumber \tilde{G}(\kp, \kp', \ell') &=& \int d\xp' e^{i(\kp'-\kp)\xp'} b_{\g}\\ &\times& \sqrt{P\left(\sqrt{\kp^2 + \frac{\ell'^2}{\chi^2}}, z(\xp')\right)}\, ,
    \label{eq:gwindow}
\end{eqnarray}
for the Line, Background and Galaxy observables, respectively. These window functions can be calculated numerically as the Fourier transform of bias, intensity, and now additionally the square root of the non-linear MPS, which we compute via CAMB\footnote{\url{https://camb.info/}}. With these expressions, the overdensity of observable X now has the simple form
\begin{align}
    \delta X(\kp', \ell') = \int \frac{d \kp}{2\pi\chi^2} \tilde{X}(\kp, \kp', \ell')\, .
\end{align}
Using this we can simplify Eq.\ref{eq:windowcov} as
\begin{align}
    \VEV{\delta X(\kpa', \ellv'_1) \delta Y^*(\kpb', \ellv'_2)} &= (2\pi)^2 \int \frac{d\kp}{2\pi\chi^2} \tilde{X}\tilde{Y}^* \delta^2{(\ellv'_1-\ellv'_2)}\, ,
    \label{eq:windowcov2}
\end{align}
where again we have absorbed the MPS entirely into the window functions themselves. Equating Eqs. \ref{eq:3dangle} and \ref{eq:windowcov2}, the angular power spectrum can be expressed as
\begin{align}
    C^{X\times Y}(\kp', \ell') = \frac{1}{\chi^2 \mathcal{L}'}\int \frac{d\kp}{2\pi} \tilde{X}(\kp, \kp', \ell')\tilde{Y}^*(\kp, \kp', \ell')\, .
    \label{eq:angleps}
\end{align}
Here we have made a choice to consider only the diagonal components of the parallel Fourier modes, that is, for $\kpa' = \kpb'$ between the overdensities. In practice, there will be physical information contained in the off diagonal covariance terms. However, promoting the formalism to include the off diagonal components quadratically increases the amount of computation (for $N$ $\kp'$ bins, from $N$ to $N(N+1)/2$ power spectra needed). Thus, in this project we only consider the diagonal terms as a quick test for our proposed joint analysis, keeping in mind that the addition of off-diagonal components may be useful for future work and surveys. 

\section{Input Models} \label{sec:models}
To test our power spectrum model and its SFR constraining capabilities, we adopt the SFR halo model presented in \citet{maniyar21}, hereafter M21. We first provide an overview of this formalism, and subsequently outline our choice of the CIB and $\cii$ intensity models in relation to the SFR. 

\subsection{SFR Halo Model}
The halo model formulates the SFR in terms of an efficiency function, $\eta$. At a halo mass $M_h$ and a redshift $z$, the star forming efficiency is given as a lognormal\footnote{We use $\log$ to signify common log of base 10. Natural log of base $e$ are denoted by $\ln$, such as in Eq. \ref{eq:eta}. Additionally, we choose to keep the base 10 parameterization of $\log{M_{\rm max}}$ from M21.} distribution over halo masses
\begin{align}
    \eta(M_h, z) = \eta_{\rm max} \times \exp{\left[-\frac{(\ln{M_h} - \ln{M_{\rm max}})^2}{2\sigma(z)^2}\right]}\, ,
    \label{eq:eta}
\end{align}
where $\sigma$ is a function of redshift that controlls the spread of the distribution
\begin{align}
    \sigma(z) = \sigma_{M_{h0}} - \tau \times {\rm max}(0, z-z_c)\, ,
    \label{eq:sigma}
\end{align}
with $z_c = 1.5$ a fixed value, signifying the maximum redshift below which $\sigma(z)$ is allowed to evolve. In total, Eq. \ref{eq:eta} is controlled by four free parameters
\begin{align}
    p_\alpha = \{\eta_{\rm max}, \log{M_{\rm max}}, \sigma_{M_{h0}}, \tau\} \, .
    \label{eq:sfrparams}
\end{align}
The SFR is subsequently defined as
\begin{align}
    {\rm SFR}(M_h, z) = \eta(M_h, z) \times {\rm BAR}(M_h, z)\, ,
    \label{eq:SFR}
\end{align}
where BAR is the baryonic accretion rate. Physically speaking, Eq. \ref{eq:SFR} describes star formation as dependent on two processes: the BAR factor which describes the rate of baryonic gas accretion by the host dark matter halo, and the $\eta$ factor that governs the efficiency of star formation from the accreted baryonic gases. The BAR is defined as
\begin{align}
    {\rm BAR}(M_h, z) = \VEV{\dot{M}(M_h, z)} \times \Omega_b(z)/\Omega_m(z)\, ,
    \label{eq:BAR}
\end{align}
where $\dot{M}(M_h, z)$ is the mass growth rate, for which M21 adopt the mean estimate from \cite{fakhouri10} with the form
\begin{align}
    \nonumber \VEV{\dot{M}(M_h, z)} &= (46.1\ \myr) \times \left(\frac{M_h}{10^{12} M_{\odot}}\right)^{1.1}\\
    &\times (1+1.11z)\sqrt{\Omega_m(1+z)^3 + \Omega_{\Lambda}}\, .
    \label{eq:mdot}
\end{align}
In Eq. \ref{eq:BAR} $\Omega_b$ and $\Omega_m$ are dimensionless cosmological parameters, representing the baryon density and total matter density, respectively. A ratio of the two is applied to filter out only the baryon part of matter accretion. It should be noted that $\Omega_m$ and $\Omega_{\Lambda}$(the dark energy density) that appear in Eq. \ref{eq:mdot} are their respective values at $z=0$ unlike those of Eq.~\ref{eq:BAR}.

For all subsequent calculations dependent on the SFR, as well as input to Fisher analysis, we use the best fit values of the SFR parameters given in Table. 1 of M21. These values are $\eta_{\rm max} = 0.42$, $ \log{M_{\rm max}}=12.94$, $\sigma_{M_{h0}}=1.75$, and $ \tau = 1.17$.

\subsection{CIB Intensity \& Bias}
We express the CIB intensity in terms of the comoving emissivity function, $j$, and integrate the quantity over redshift in the form
\begin{align}
    I_{\rm CIB}(\lambda) = \int dz \frac{d\chi}{dz} a j(\lambda, z)\, ,
    \label{eq:ICIB}
\end{align}
where $\chi$ is the comoving distance and $a = \frac{1}{1+z}$ is the scale factor of the universe. As discussed before, for the CIB fluctuation defined in Eq. \ref{deltaCIB}, we are interested in the redshift derivative of the CIB intensity. From \ref{eq:ICIB} we simply have:
\begin{align}
    \frac{dI_{\rm CIB}(\lambda, z)}{dz} = \frac{d\chi}{dz} a j(\lambda, z)\, .
\end{align}
Our definition of the CIB emissivity is also adapted from M21. The definition begins with the specific emissivity given by:
\begin{eqnarray}
    \nonumber \frac{dj}{d\log{M_h}}(\lambda, M_h, z) &=& \frac{dn}{d\log{M_h}}(M_h, z) \chi^2(1+z) \\
    &\times& \frac{{\rm SFR}(M_h, z)}{K} S^{\rm eff}(\lambda, z)\, ,
    \label{eq:djdm}
\end{eqnarray}
where $\frac{dn}{d\log{M_h}}$ is the halo mass function calculated from \citet{2008ApJ...688..709T}, $K=1\times 10^{-10} \myr L_{\odot}^{-1}$ is the Kennicutt constant for a Chabrier IMF \citep{2003PASP..115..763C} which scales the infrared luminosity-SFR ratio, and $S^{\rm eff}$ is the effective SED. The emissivity can then be obtained from Eq. \ref{eq:djdm} by performing an integral over halo masses.

Note that M21 introduce a formalism that models both host and satellite dark matter halos (or subhalos) to study both correlations within one particular halo (denoted $P_{1h}$), and correlations between halos on large scales (denoted $P_{2h}$). In our analysis we focus on the large angular scales and thus only consider on the 2-halo term ($P_{2h}$). In principle, however, subhalos effects can be easily added by modifying the emissivity formalism above, albeit at a non-negligible increase to computation time due to calculations of the subhalo SFR. 

The effective bias for the CIB can be obtained as follows. As the CIB intensity is halo model based, the most natural choice for the bias term is the halo bias, which we calculate based on \citet{2010ApJ...724..878T}. In its natural form, the halo bias is dependent on both halo mass and redshift, whereas the final form of the CIB intensity times bias should only be redshift dependent. Thus, we introduce the halo bias to the definition of the specific emissivity of Eq. \ref{eq:djdm} to integrate out the mass dependence on the bias, namely:
\begin{align}
    b_{\rm CIB}(z)j(\lambda, z) &= \int d \log{M_h}\ b(M_h, z)\frac{dj}{d\log{M_h}}(\lambda, M_h, z)\, ,
    \label{eq:CIBbias}
\end{align}
where $b(M_h, z)$ on the RHS is the halo bias. 

\subsection{$\cii$ Intensity \& Bias}
Since early observations of the $\cii$ line in star-forming galaxies, numerous models have emerged in literature to accurately model the $\cii$ luminosity-SFR relation. Many existing models parameterize a $L_{\cii}$-SFR power law relation of the form:
\begin{align}
    \log{L_{\cii}} = \alpha\log{\rm SFR} + \beta \, ,
    \label{eq:lcii}
\end{align}
where $L_{\cii}$ is measured in solar luminosity units, $L_{\odot}$, and $\alpha$, $\beta$ are the free parameters of the power law. The values of these parameters have been fitted to a combination of observations and simulations, and can vary quite dramatically depending on redshifts and assumptions of the SFR (metalicity dependence, etc.). In this analysis, we use the empirical relation derived by the mock light cone simulation analysis from \cite{yang21}, with values of $\alpha=1.26$ and $\beta=7.1$ \footnote{These values were derived from the \cite{yang21} semi-analytical model, but were not included as part of their publication, and were given to us via private communication}. As the EXCLAIM $\cii$ observations is also a fiducial experiment considered in their simulations, these values become a natural choice for our subsequent tests based on EXCLAIM specifics. 

To  obtain the $\cii$ intensity from its luminosity, we follow the formalism presented in \cite{bernal19}. A few steps of this conversion are as follows. Given an emission line with rest frame frequency $\nu$, the intensity can be written as
\begin{align}
    I(z) = \frac{c}{4\pi\nu H(z)} \rho_L(z)\, ,
\end{align}
where $c$ is the speed of light, $H(z)$ is the Hubble expansion rate, and $\rho_L$ is the luminosity density function, whose average is given by
\begin{align}
    \VEV{\rho_L(z)} = \int dM_h \frac{dn}{dM_h} L(M_h, z)\, ,
\end{align}
an integral of the luminosity function  over halo masses, weighted by the halo mass function $\frac{dn}{dM_h}$. 

\cite{bernal19} also provide a framework for the effective bias for the $\cii$ line. Similar to the CIB effective bias, the $\cii$ line that traces galaxies within halos is sensitive initially to the halo bias. To linear order, the effective bias is then obtained via a weighted average of the halo bias over the $\cii$ luminosity
\begin{align}
    b_{\cii}(z) = \frac{\int dM_h\ L(M_h, z) b(M_h, z) \frac{dn}{dM_h}(M_h, z)}{\int dM_h\ L(M_h, z)\frac{dn}{dM_h}(M_h, z)}\, .
    \label{eq:ciibias}
\end{align}

\section{Power Spectra Results} \label{sec:powerspec}
By applying the intensity and bias models of Sec.~\ref{sec:models} to the power spectra formalism of Sec.~\ref{sec:theory}, we numerically compute the joint line and continuum intensity-cross-galaxy power spectra. As the fiducial experiment, we choose relevant survey parameters of the EXCLAIM survey as inputs to the power spectrum model. We will first briefly discuss the impact that various parameters have on the computation of the power spectra.

The EXCLAIM telescope is tuned to a wavelength range of $553\micron \leq \lambda \leq 711\micron$. For the $\cii$ line with rest frame wavelength of $158 \micron$, this corresponds to a redshift window of $2.5 \leq z \leq 3.5$. Converting this redshift window into comoving distances, this sets the bounds for all line-of-sight integrals that concern the line intensity fluctuations and the galaxy overdensities, such as those in Eqs. \ref{linefourier} and \ref{eq:gfourier}. In addition, this boundary fixes the value for $\mathcal{L}'$, the LOS window size which appears in the angular power spectrum expressions \ref{eq:3dangle} and \ref{eq:angleps}, at $\chi(z=3.5) - \chi(z=2.5) = 984.56 {\rm Mpc}$. 

With this relatively small window along the LOS, we find that the product of the $\cii$ intensity and effective bias follows a linear relation in redshift. Thus, as an approximation, we treat the $\cii$ intensity and bias both as constants over redshift, which can be extracted from the $\cii$ window function. Their fixed values are simply taken to be their values at the center of the redshift window, $z=3$, which is also the average value within this window. We can therefore modify Eq. \ref{eq:lwindow} as
\begin{align}
    \tilde{L}(\kp, \kp', \ell') = \frac{b_{\cii}^{z=3} I_{\cii}^{z=3}}{b_{\g}} \tilde{G}(\kp, \kp', \ell'), ,
    \label{eq:deltaline2}
\end{align}
where $\tilde{G}$ is the galaxy window function given by Eq. \ref{eq:gwindow}. In this way, the line and galaxy window functions share the identical LOS integral. This approximation drastically reduces the amount of required computation, as both the line and galaxy window functions can be obtained by simply scaling the result of the LOS integral of Eq. \ref{eq:deltaline2} by the appropriate constant factors.

\begin{figure*}
    \centering
    \includegraphics[width=0.7\linewidth]{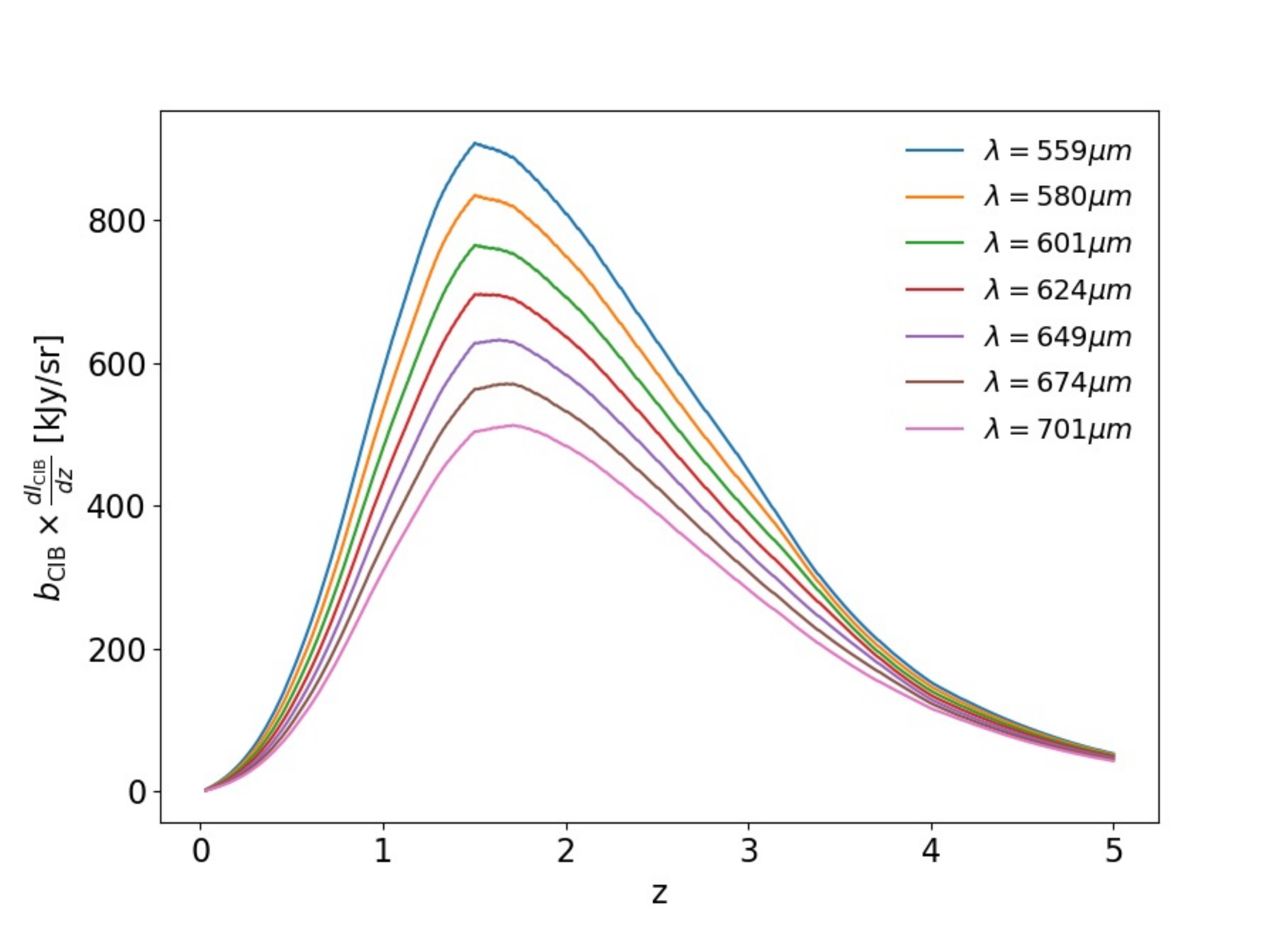}
    \caption{The CIB intensity distribution over redshifts weighted by the halo mass integrated effective bias, at select wavelengths within the EXCLAIM detection range.}
    \label{fig:ICIB}
\end{figure*}

On the other hand, CIB contribution at a given wavelength comes from a very broad range of redshifts. Within this range the CIB intensity can vary quite considerably. As shown in Fig. \ref{fig:ICIB}, approximating the CIB intensity at the redshift of its average value is far from ideal, due to its much wider and non-linear evolution in redshift compared to the $\cii$. In addition, the CIB intensity at different wavelengths reveal that the peak of the distribution is wavelength dependent, signifying that we cannot fix one redshift to sample the CIB for all wavelengths without further loss of information. Thus in the CIB window function we maintain the intensity and bias terms as redshift dependent quantities, and numerically integrate them along with the MPS. Here the large redshift span makes numerical integration costly; to alleviate the computational burden, we notice that the CIB intensity drops quite significantly beyond $z > 5$, a feature that is consistent across all wavelengths, as can be seen in Fig. \ref{fig:ICIB}. We therefore set the upper bound of the LOS integral at a cutoff redshift of $z=5$, discarding the less significant high redshift portion, effectively reducing the computational cost of each integral by half.

Although the ultimate goal is to cross-correlate a galaxy field with the joint intensities of both the CIB and the $\cii$ line, this operation is identical to cross-correlating each intensity with galaxies separately, and combining the results in power spectrum space. Namely, the computation can be done via the following distributive identity
\begin{align}
    \VEV{(\delta I_{\rm CIB} + \delta I_{\cii})\ \delta \g^*} = \VEV{\delta I_{\rm CIB}\ \delta \g^*} + \VEV{\delta I_{\cii}\ \delta \g^*}\, .
    \label{eq:cross}
\end{align}

We note here that the CIB part of the cross-correlation will be a complex quantity. This is due to both the different redshift window size of the CIB and galaxy fields, as well as due to the content of their window functions (see Eqs. \ref{eq:bwindow} and \ref{eq:gwindow}, respectively). Whereas the only redshift varying quantity within the galaxy window function is the MPS, its CIB counterpart contains additional factors of intensity and bias that both evolve with redshift. Thus, when cross-correlating the two quantities using Eq. \ref{eq:cross}, the product of the window functions will introduce complex cross terms. On the other hand, the $\cii$-galaxy cross-power spectrum is purely real assuming the approximation of Eq. $\ref{eq:deltaline2}$, in which all biases and intensities are real, constant values, and the $\cii$ intensity shares the identical complex window function with the galaxy overdensity field. 

Physically speaking, this effect is once again a result of the redshift distribution of the intensities. In particular, as the $\cii$ line shares comparable redshift ranges as the galaxy field, the two fields trace the same underlying matter and will produce real-valued cross-correlation. Conversely, while being in the same wavelength window as the $\cii$ intensity, the CIB traces matter over a much larger distance scale compared to the galaxy field, and the cross-correlation of the matter overdensity over redshift slices with such large separations gives rise to the complex valued cross-correlation. 

\begin{figure*}
    \centering
    \includegraphics[width=\linewidth]{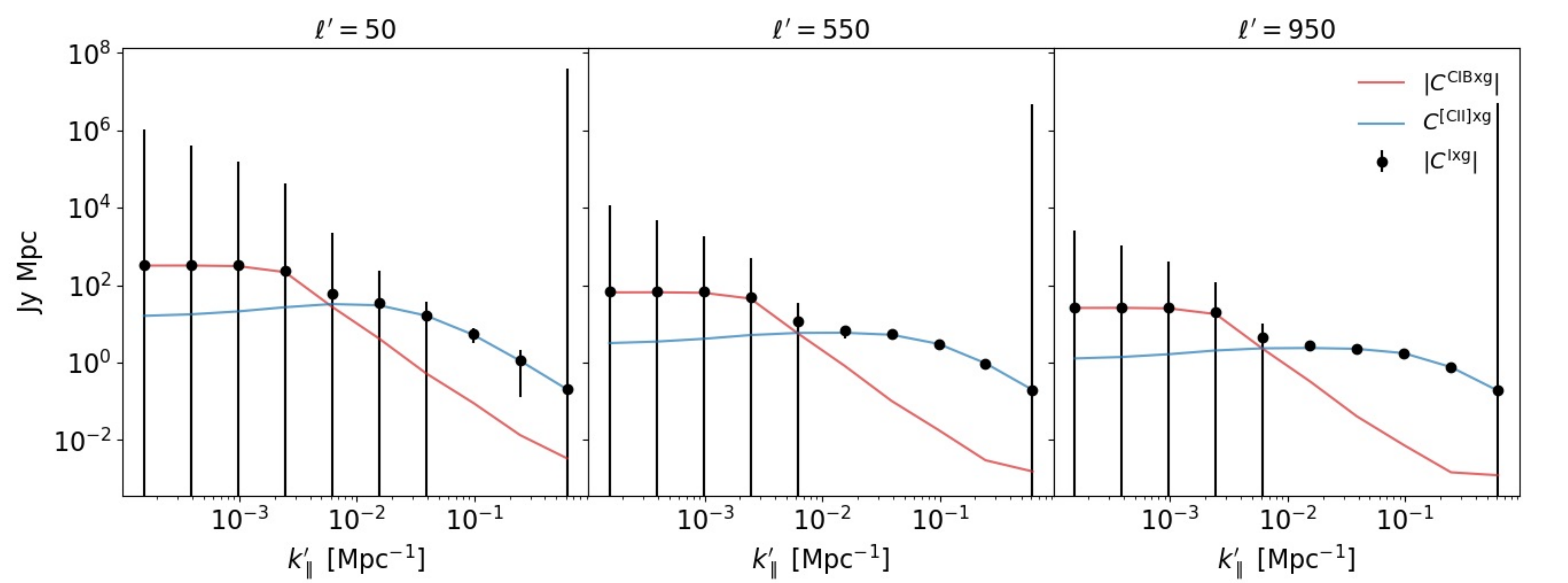}
    \caption{Angular power spectra results for cross-correlation of line and continuum intensities with the eBOSS galaxy field, calculated via Eq. \ref{eq:angleps} and the intensity and galaxy window functions. The three panels (left to right) show results corresponding to $\ell' = 50$, $550$, and $950$. In each panel the cross power spectrum of the total intensity is shown in black, with cosmic variance and survey noise error bars according to Eq. \ref{eq:inputvar}. For explicit comparison over $\kp'$, shown in red and blue are the CIB cross galaxy and $\cii$ cross galaxy power spectra, respectively, which combine to form the total power in black.}
    \label{fig:Cl}
\end{figure*}

In Fig. \ref{fig:Cl} we show the power spectra results over $\kp'$ for select values of $\ell'$. In the same figure we show the binned cross-power spectra of CIB cross galaxy and $\cii$ cross galaxy separately to observe their respective behaviors over $\kp'$ bins. These intensity components behave as expected,  where the total power is dominated by the continuum emission at low $\kp'$ bins, which quickly decays with higher $\kp'$ values as the line emission becomes dominant. The error on the joint intensity (black error bars) include both effects of cosmic variance and expected noise of all relevant observables. For intensity cross galaxy, this entails the detector noise of the EXCLAIM survey, $P_N$, and the shot noise of the galaxy catalogue, $1/\bar{n}_{\g}$, with $\bar{n}_{\g}$ the galaxy number density. We model this noise variance with the following form
\begin{align}
    \nonumber \sigma^2(\kp', \ell') &= \left|C^{\rm I\times \g}(\kp', \ell')\right|^2 + \left[C^{\rm I\times I}(\kp', \ell') + \frac{P_N}{\chi^2 W(\kp', \ell')}\right] \\
    &\times \left[C^{\rm \g \times \g}(\kp', \ell') + \frac{1}{\chi^2 \bar{n}_{\g}}\right] \, ,
    \label{eq:inputvar}
\end{align}
where each noise component receives a factor of $1/\chi^2$ due to the flat-sky approximation, and the intensity auto-power spectrum, $C^{\rm I\times I}$, can be expanded to yield
\begin{align}
    C^{\rm I \times I} = C^{\rm CIB\times CIB} + 2\ {\rm  Re}\left[C^{\rm CIB\times \cii}\right] + C^{\rm \cii \times \cii}\, ,
\end{align}
since the correlation amongst the intensities has the form:
\begin{align}
   \nonumber \VEV{\delta I_{\rm Tot} \delta I_{\rm Tot}^*} &= \VEV{(\delta I_{\rm CIB} + \delta I_{\cii})\ (\delta I_{\rm CIB} + \delta I_{\cii})^*}\\
   \nonumber &= \VEV{\delta I_{\rm CIB} \delta I_{\rm CIB}^*} + \VEV{\delta I_{\rm CIB} \delta I_{\cii}^*}\\ 
   \nonumber &+ \VEV{\delta I_{\cii} \delta I_{\rm CIB}^*} + \VEV{\delta I_{\cii} \delta I_{\cii}^*}\\
   \nonumber &= \VEV{\delta I_{\rm CIB} \delta I_{\rm CIB}^*} + \VEV{\delta I_{\cii} \delta I_{\cii}^*} \\
   &+ 2\ {\rm Re}\left[\VEV{\delta I_{\rm CIB} \delta I_{\cii}^*}\right]\, ,
\end{align}
where the cross terms in the second line are complex conjugates of the same quantity, which are combined to yield only its real part in line three. Additionally in Eq. \ref{eq:inputvar}, $W(\kp', \ell')$ is a Gaussian window function of the form
\begin{align}
   W(\kp',\ell')=e^{-\delta \xp'^2\kp'^2-\theta_{\rm res}^2\ell'^2}\, ,
\end{align}
with $\theta_{\rm res}$ the angular resolution of the survey, and $\delta \xp'$ the channel width resolution along the LOS given by
\begin{align}
    \delta \xp = \frac{c(1+z)}{H(z)R}\, ,
\end{align}
where finally R is the survey's spectral resolution. The values of these parameters are provided by the EXCLAIM forecast paper \citep{forecast}.

\section{Fisher forecast results and discussions}\label{sec:fisher}

We move on to deriving the final SFR constraint forecasts that our power spectrum model produces. To do so we rely on the utility of the Fisher formalism, and compute a Fisher matrix that encodes the covariances between the input SFR parameters. We will first outline the exact formalism that is appropriate for our model.

\subsection{Fisher Formalism}
Via the principles of Bayesian analysis, the Fisher formalism transforms uncertainties of the model dependent observables onto the uncertainties of the model itself. In general one can construct a Fisher matrix as the following
\begin{align}
    F_{\alpha\beta} = \sum_{i=1}^N \frac{1}{\sigma_i^2}\frac{\partial f_i}{\partial p_\alpha}\frac{\partial f_i}{\partial p_\beta}\, ,
    \label{eq:sumFisher}
\end{align}
where $F_{\alpha\beta}$ is the Fisher matrix, $p_{\alpha}$ are the model parameters, $N$ is the number of available observables and $\sigma_i^2$ is the expected variance on the observable $f_i$. It can be shown that the inverse of the Fisher matrix is:
\begin{align}
    F_{\alpha\beta}^{-1} = \VEV{\delta p_\alpha \delta p_\beta}\, ,
    \label{eq:inverse}
\end{align}
which is precisely the covariance matrix, $\rm{Cov}_{\alpha \beta}$.

For this analysis, the parameters $p_\alpha$ are the SFR halo model parameters of Eq. \ref{eq:sfrparams}, and the observables $f_i$ are the intensity cross galaxy power spectra from Sec.~\ref{sec:powerspec}. We assume that each value of the power spectra at given wavemode $\kp'$ and multipole $\ell'$ is an independent observation. Since the power spectra are continuous along these two axes, it is more appropriate to perform an averaging integral along both dimensions in place of the sum in Eq. \ref{eq:sumFisher}. With this replacement the Fisher matrix has the form
\begin{align}
    F_{\alpha\beta} = 4 \pi f_{\rm sky} \mathcal{L'} \int \frac{d\kp' d^2 \ell'}{(2\pi)^3} \frac{1}{\sigma^2} {\rm Re}\left[\left(\frac{\partial C^{\rm I\times g}}{\partial p_\alpha}\right)\left(\frac{\partial C^{\rm I\times g}}{\partial p_\beta}\right)^*\right]\, ,
    \label{eq:fisher}
\end{align}
where the observable variance, $\sigma^2$ is just that of the power spectra variance given by Eq. \ref{eq:inputvar}. The integral is averaged by the survey volume, $4 \pi f_{\rm sky} \mathcal{L'}$, where $f_{\rm sky}$ is the percentage of the full sky covered by the survey instrument. 

We compute two sets of Fisher matrices to evaluate the additional constraining effect of introducing the CIB. For the fiducial calculation that simulates the conventional approach for LIM surveys, we input only the $\cii$ part of the intensity (i.e. $C^{\rm I\times g} = C^{\cii \times {\rm g}}$). Additionally, we place a lower cutoff along $\kp'$, where signals below this cutoff are discarded due the considerable continuum bias within these scales. We set this lower bound at $\kp' = 0.017 {\rm Mpc}^{-1}$, a scale at which the CIB power only constitutes 10\% of the total power spectrum, which only minimally biases constraints. In contrast to this treatment, we compute the joint signal Fisher matrix where the observable contains both line and CIB signals (i.e. $C^{\rm I\times g} = C^{\cii \times {\rm g}} + C^{\rm CIB \times g}$). In this case, we no longer impose the $\kp'$ lower bound in this result as the dominating CIB at low $\kp'$ bins is now treated as signal rather than a source of bias.

Obtaining the covariance matrix according to Eq. \ref{eq:inverse} naturally requires an invertible Fisher matrix. However, for the first procedure where only the $\cii$ intensity is used as input, the resulting Fisher matrix is highly degenerate. This is unsurprising as the $\cii$ is evaluated at only one value of $b_{\cii}I_{\cii}$, while the parameter space of the SFR model is of dimension 4. We circumvent this problem by adding a prior Fisher which forecasts the same parameters, using the additive property of Fisher matrices. This prior is the Fisher forecast given by M21 using analyses of constraining the star formation rate density (SFRD). Since the prior and $\cii$ Fishers are derived from independent signals, the overall result becomes invertible. The $\cii$ only SFR constraint is then
\begin{align}
    \cov^{\cii} = \left(F^{\rm prior} + F^{\cii}\right)^{-1}\, ,
    \label{eq:ciifisher}
\end{align}
where $\cov^{\cii}$ is the covariance matrix, and again $F^{\cii}$ is computed using Eq. \ref{eq:fisher} using only $\cii$ cross galaxy power spectra as input. Conversely, while the $\rm CIB+\cii$ cross galaxy joint Fisher is invertible by default, we also combine this result with the halo prior such that the two cases are consistent for a meaningful comparison.

\subsection{Constraint Results}
\begin{figure*}
    \centering
    \includegraphics[width=0.7\linewidth]{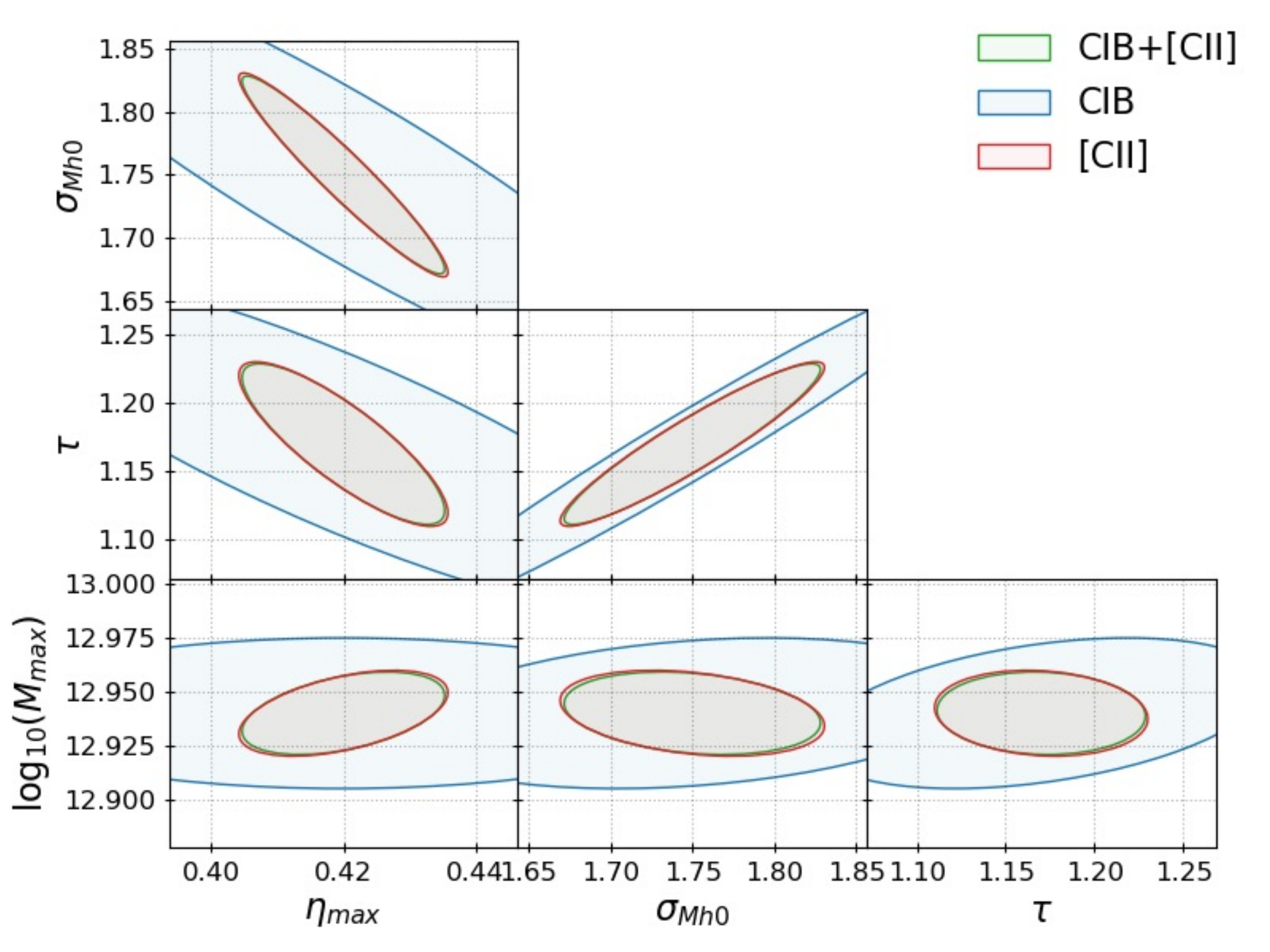}
    \caption{Fisher constraints on SFR halo model parameters. We use the current EXCLAIM survey noise and sky coverage parameters, including an $f_{\rm sky}$ of 0.007. In each panel, we show the joint continuum and line intensity constraints in green. In blue and red are the separate constraints using CIB and $\cii$ intensities independently. In addition, all three intensity Fisher matrices are first combined with the halo prior Fisher, due to the $\cii$ Fisher being non-invertible.}
    \label{fig:current}
\end{figure*}

In Fig. \ref{fig:current} we show the joint CIB and $\cii$ cross galaxy constraints for the SFR halo model parameters, assuming EXCLAIM as the fiducial survey. These results show that the addition of the CIB will improve the constraint only minimally. This marginal increase argues in favor of the much simpler treatment of cutting the CIB in power spectrum space, in light of the computational difficulty in incorporating the CIB.

While it is clear that a joint analysis with the CIB will not significantly benefit the EXCLAIM survey, we can inquire as to the possible improvements that a joint analysis can bring to future EXCLAIM like experiments, where survey parameters can be expected to receive considerable improvements. In particular, future detectors are expected to cover a wider transverse plane; equivalently, this corresponds to an increase in the survey sky coverage, or in numerical terms an increase to the  $f_{\rm sky}$ parameter. The value of the $f_{\rm sky}$ is especially significant for the Fisher forecast, as the Fisher matrix is linearly proportional this quantity, as shown in Eq. \ref{eq:fisher}. Thus, any significant boost to the $f_{\rm sky}$ should correspondingly result in an overall increase in the Fisher magnitude, or equivalently, a reduction in the parameter covariance, being the inverse of the Fisher matrix. Experiments capable of probing a larger sky region will therefore achieve superior constraints, and it is possible that in this framework, the benefit of adding the CIB will emerge as non-negligible, and even possibly significant. 

\begin{figure*}
    \centering
    \includegraphics[width=0.7\linewidth]{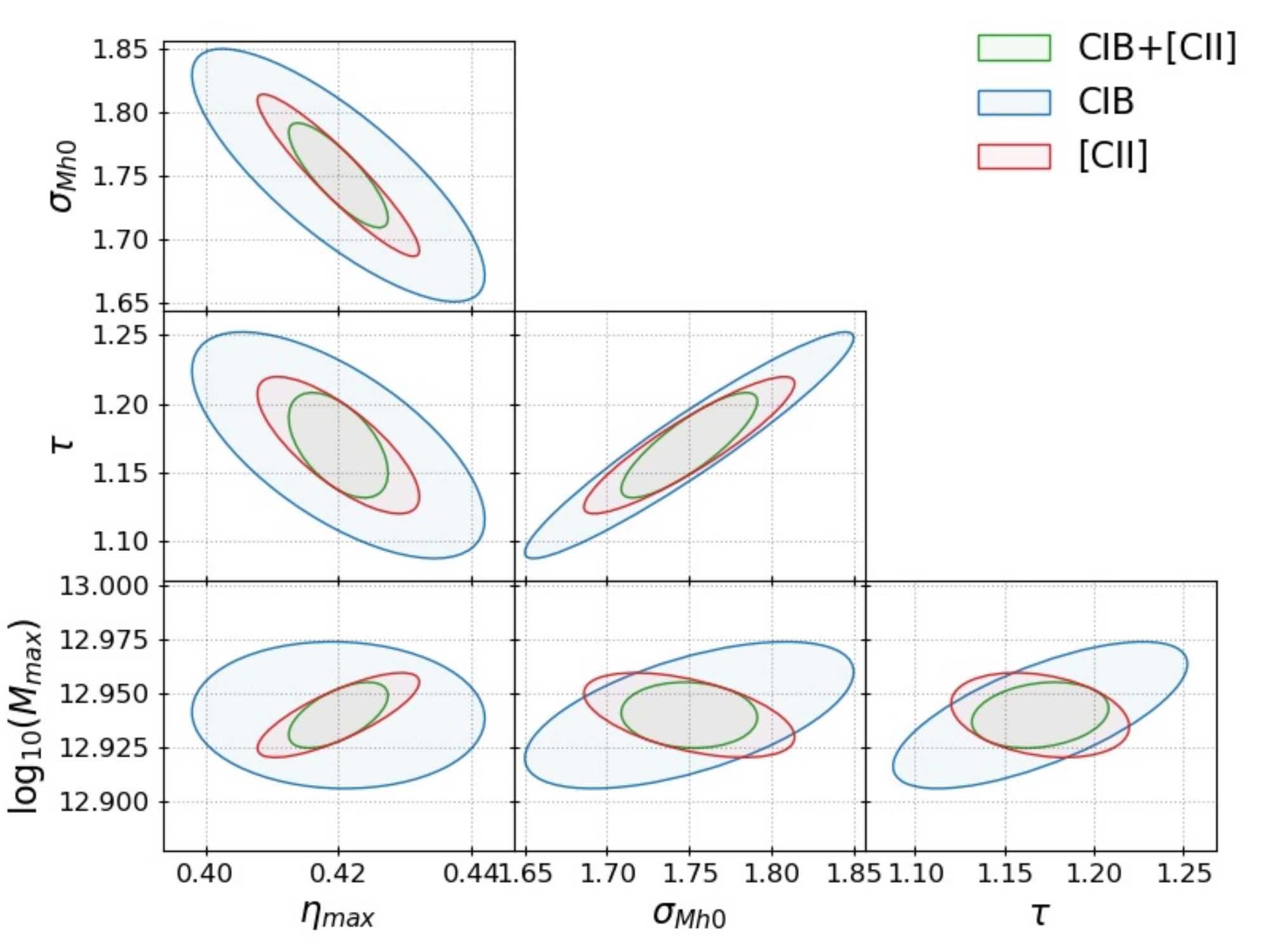}
    \caption{Fisher constraints after an optimistic boost to the survey $f_{\rm sky}$ has been applied. All details of the figure are identical to those in Fig. \ref{fig:current}, but results are calculated using $f_{\rm sky} = 0.1$, roughly 15 times the current EXCLAIM value of $f_{\rm sky} = 0.007$.}
    \label{fig:large_current}
\end{figure*}

We test this possibility by repeating our calculations and producing a new set of survey parameters, where we assume a value of $f_{\rm sky}^{\rm future} = 0.1$, a large increase from the current EXCLAIM value of $f_{\rm sky}^{\rm current} = 0.007$. With all other parameters kept identical to previous calculations, this new set of constraint forecasts are shown in Fig. \ref{fig:large_current}.

We observe that the CIB contribution begins to have promising effects. At best, we observe a $40\%$ reduction in the parameter variance when comparing the constraints before and after the CIB has been combined with the $\cii$ intensity. Additionally, the individual contours within Figs. \ref{fig:current} and \ref{fig:large_current} reveal that the seaparate CIB cross galaxy and $\cii$ cross galaxy constraints appear to improve disproportionally with increasing $f_{\rm sky}$; that is, the CIB cross galaxy constraint contours shrink more rapidly than those of the $\cii$ cross galaxy, even though the same $f_{\rm sky}$ increase is applied to both approaches. Consequently, improvements that are evident in the joint constraint contours are predominantly driven by improvements in the CIB based constraints, as the $\cii$ cross galaxy contours (red contours in Figs. \ref{fig:current} and \ref{fig:large_current}) are shown to shrink minimally across the constraints with two values of the $f_{\rm sky}$. 

\begin{figure*}
    \centering
    \includegraphics[width=0.7\linewidth]{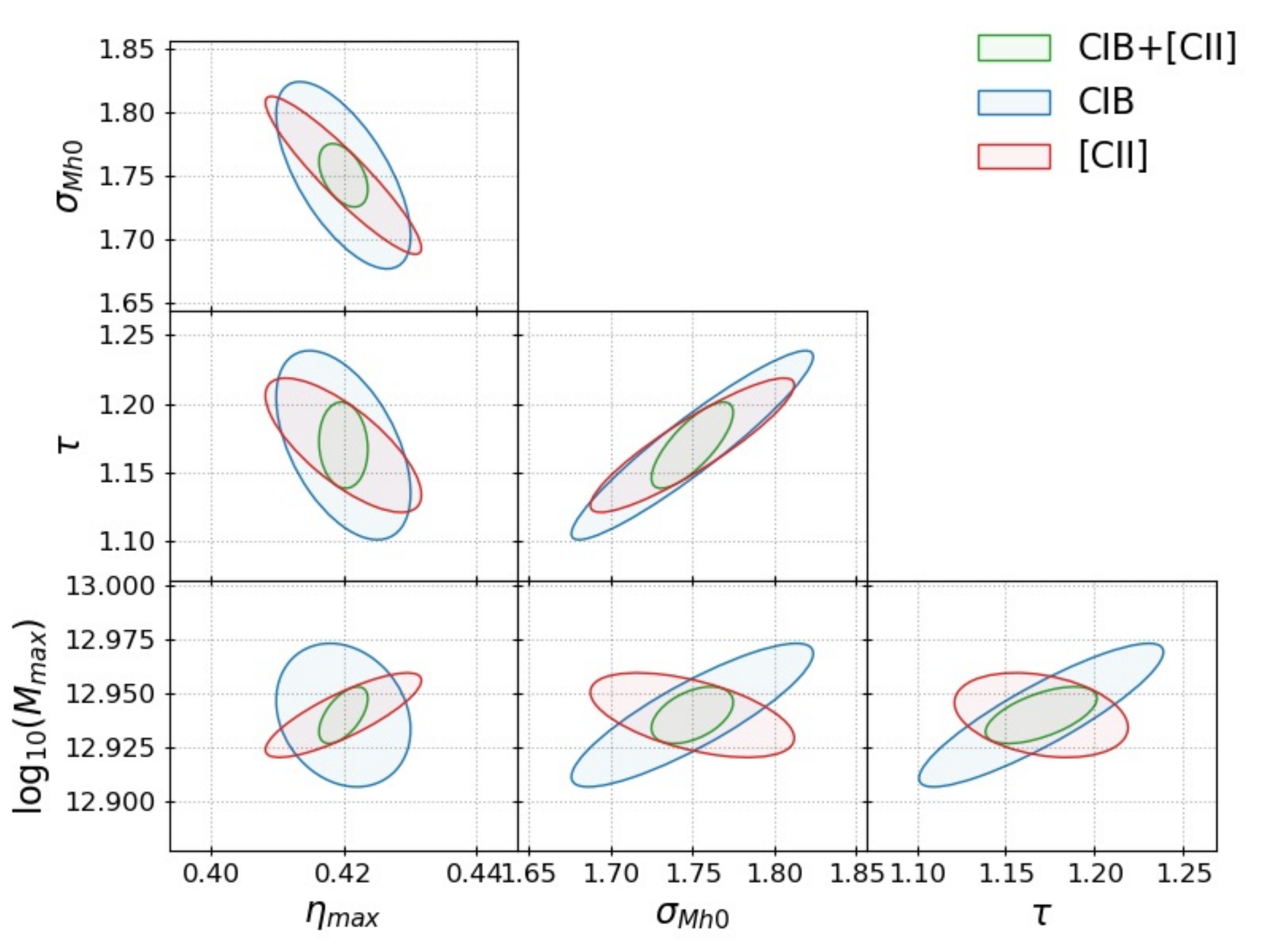}
    \caption{More optimistic Fisher constraints achievable by future surveys. Inheriting the large $f_{\rm sky}$ value of 0.1 in Fig. \ref{fig:large_current}, these constraints come from additionally reducing all noise parameters in the input variance by a factor of 10.}
    \label{fig:large_reduce}
\end{figure*}

This effect becomes even more apparent when we are also optimistic about the noise levels of the future surveys. We produce a third set of results, where in addition to a boosted $f_{\rm sky}$ value of 0.1, we also reduce the detector noise and the galaxy shot noise parameters in Eq. \ref{eq:inputvar} each by a factor of 10 from the current EXCLAIM values. These constraints are shown in Fig. \ref{fig:large_reduce}

With improved noise levels, the CIB constraints continue to show noticeable reductions. On the other hand, we notice that the $\cii$ constraints are nearly non-responsive to reduced survey noises, as can be seen in Figs. \ref{fig:large_current} and \ref{fig:large_reduce}. We can reasonably conclude that as we tune particular survey parameters - $f_{\rm sky}$ and noises in our analysis - towards optimistic future values, the advantage we observe in the joint continuum and line intensity constraints become significant, and are dominated largely (or even entirely) by the responsiveness of the CIB based constraints to these survey improvements.

\subsection{Discussion of the CIB Advantages}
These sets of forecast results argue for a joint CIB + line emission data analysis for future surveys, and it is worthwhile to understand how it is that the CIB is capable of improving the constraints at these scales. To do so, first we must discuss the observation that the $\cii$ cross galaxy constraints appear to reach a "minimum" with improving survey specifics, and cease to be responsive to any change in $f_{\rm sky}$ increase or the survey noise reductions. This can be somewhat counter-intuitive, but turns out to be key in understanding the strength of introducing the CIB. 

As mentioned before, when computing the parameter covariance from using only the $\cii$ intensity, it is necessary to add a prior Fisher matrix to break the degeneracy of the $\cii$ cross galaxy Fisher on its own, as given by Eq. \ref{eq:ciifisher}. The prior as previously discussed is adapted from the studies of M21, and thus a fixed quantity and not sensitive to any changes in the $\cii$ cross galaxy Fisher. Then, as we tune the survey parameters that affect the $\cii$ Fisher information, only the second term in Eq. \ref{eq:ciifisher} will change. Then, suppose we increase the value of the $f_{\rm sky}$ by a multiplicative factor of $\alpha$; since the Fisher matrix is linearly dependent on $f_{\rm sky}$, the new covariance after this $f_{\rm sky}$ increase is simply
\begin{align}
    \cov^{\cii} = \left(F^{\rm prior} + \alpha F^{\cii}\right)^{-1}\, .
    \label{eq:SMleft}
\end{align}
If $F^{\cii}$ were invertible on its own, increasing the value of $\alpha$ indefinitely would simply let the overall inverse be dominated by $(F^{\cii})^{-1}$. In reality, as $F^{\cii}$ is highly degenerate, it cannot be the dominating contribution in the overall covariance, even if the $\alpha$ factor increases without bounds. To probe the effect with this consideration, we employ the Sherman Morrison formula \citep{sm50}, which is an exact analytical expansion of the inverse of sum of matrices, so long as the matrices satisfy certain conditions, which we will outline here. Given an invertible matrix $A$ and a rank 1 matrix $B$, and given that $A+B$ is invertible, the formula reads:
\begin{align}
    (A+B)^{-1} = A^{-1} - \frac{1}{1+\boldsymbol{Tr}[A^{-1}B]}A^{-1}BA^{-1}\, ,
    \label{eq:SM}
\end{align}
where $\boldsymbol{Tr}[]$ denotes the trace operator. The Fisher matrices we have happen to fit the exact requirements to use the Sherman Morrison formula: an invertible prior Fisher matrix $A$, and a rank 1 $\cii$ Fisher matrix $B$. When we apply the $\alpha$ boost of Eq. \ref{eq:SMleft}, Eq. \ref{eq:SM} becomes
\begin{align}
    \nonumber (A+(\alpha B))^{-1} &= A^{-1} - \frac{1}{1+\boldsymbol{Tr}[A^{-1}(\alpha B)]}A^{-1}(\alpha B)A^{-1}\\
    &= A^{-1} - \frac{\alpha}{1+
    \alpha\boldsymbol{Tr}[A^{-1}B]}A^{-1}BA^{-1} \, .
    \label{eq:SMboost}
\end{align}
With this expression, we may evaluate the large $\alpha$ limit and observe the resulting covariance as $\alpha$ grows indefinitely
\begin{align}
    \lim_{\alpha \longrightarrow \infty} (A+(\alpha B))^{-1} &= A^{-1} - \frac{1}{\boldsymbol{Tr}[A^{-1}B]}A^{-1}BA^{-1} \, .
    \label{eq:SMlimit}
\end{align}
As the Sherman Morrison formula is exact, the result of Eq. \ref{eq:SMlimit} shows that as we increase the $f_{\rm sky}$ of the survey, the overall covariance of Eq. \ref{eq:ciifisher} will converge to a set of finite minimum values, rather than decreasing indefinitely. This can again be attributed to the highly degenerate nature of the $\cii$ Fisher matrix, and explains the observed "bottleneck" effect on the $\cii$ constraints in Figs. \ref{fig:large_current} and \ref{fig:large_reduce}, with the $\cii$ covariance having reached its lower limit of Eq. \ref{eq:SMlimit}. 

This nature of the $\cii$ intensity gives the CIB an added benefit. Aside from being an SFR sensitive source, the addition of the CIB introduces a separate signal, thus breaking the degeneracy of the Fisher matrix. With an independently invertible Fisher matrix, any increase in the $f_{\rm sky}$ will be correspondingly met with an improvement in the constraints without the complications that combining a fixed prior and a singular matrix introduces. Effectively, not only will future detections of the CIB serve to boost the signal information on its own, but its ability to break the $\cii$ Fisher matrix's degeneracy will also allow surveys to fully reap the benefits of increased resolution and sky coverage. 

\section{Conclusions}\label{sec:conclusion}
In this paper we explore the prospects of combining the CIB continuum and the $\cii$ line intensity into one signal that simulates the reality of continuum contamination in LIM experiments. This is done in recognition of the CIB being a powerful tracer of the star formation process, similarly to the $\cii$ line, inspiring us to evaluate the SFR constraining power of the joint signals. 

To successfully combine the signals, we carefully define the respective intensity fluctuations necessary, as well as a power spectrum model for cross-correlating the intensities with a galaxy overdensity field. While the line intensity allows for a one-to-one correspondence between detected wavelength and source redshift, the continuum nature of the CIB obstructs this property. Circumventing this difficulty involves understanding the underlying SED of the CIB intensity. As LIM surveys calibrate the detector to a certain wavelength range, within which a particular line from a given redshift can be expected, we define a coordinate system also interpreted from the detected signal wavelength; for the CIB, this involves integrating over the entire redshift range of the SED at each wavelength, thus making the addition of CIB to LIM signal processing a non-trivial task.

For the SFR and relevant intensities, we choose  physically motivated models in literature that come from a combination of observation and light cone simulation analyses. The SFR model in particular from M21 is a widely applicable halo model with only four parameters, making it ideal for our analysis. Furthermore, we utilize the specifics of the eBOSS quasar catalogue to perform cross-correlations between relevant intensities and the galaxy overdensity to decorrelate the effects of dust foregrounds present in the Milky Way. We assume various fiducial survey scenarios to test the new SFR constraints achievable with the addition of the CIB data, and compare these results to identical calculations where only the $\cii$ is used. In particular, we assume the specifics of the EXCLAIM survey, a prominent upcoming LIM experiment with probing the $\cii$ line for star formation constraints as one of its main goals. 

We perform a Fisher analysis on our power spectrum results to forecast effective constraints on the SFR parameters. As is traditionally done, we recognize that our $\cii$ Fisher matrix is highly degenerate, and thus apply a prior Fisher matrix computed by M21 to extract the effective parameter covariance. Based on the current survey specifics of EXCLAIM, we find that the addition of the CIB yields only a minimal improvement over the constraints where only the $\cii$ is considered. However, we test scenarios in which we assume improved values of instrumental resolution and sky coverage, which can realistically be expected in future LIM surveys. In these regimes, we find that the addition of the CIB quickly shows its advantage, noticeably widening the gap between the $\cii$ only constraints and the joint signal constraints. We find additionally that the $\cii$ only covariance experiences minimal reductions in response to improvements in the survey parameters, which can be attributed to the singular nature of the $\cii$ Fisher matrix, made apparent by the Sherman Morrison formula. Addition of the CIB in this way is both a source of additional SFR information, and a separate signal capable of breaking the degeneracy of the line intensity Fisher matrix to allow for the expected constraints reduction from improved survey levels. While the current surveys like EXCLAIM will not benefit greatly from considering the CIB, especially in light of its complexity and difficulty in numerical computation, we conclude nonetheless that it will soon become a valuable source of signal for future LIM surveys, capable of further maximizing the constraints on the star formation process. 

\section*{Acknowledgements}
Many current and former members of the Pullen research group at NYU have provided great help to this project, with particular pointers and suggestions from Patrick Breysse, Shengqi Yang and Yucheng Zhang. Andreas Tsantilas at NYU CCPP provided helpful discussions on singular Fisher matrices. ARP was supported by NASA under award numbers 80NSSC18K1014, NNH17ZDA001N, and 80NSSC22K0666, and by the NSF under award number 2108411. ARP was also supported by the Simons Foundation.



\bibliographystyle{mnras}
\bibliography{example} 





\bsp	
\label{lastpage}
\end{document}